\documentclass[10pt]{mpi08} \usepackage{graphicx} 
\usepackage{cite,./mcite}  
\begin{document} \title{Recent Progress in Jet Algorithms 
and Their Impact in Underlying Event Studies
\footnote{~Talk given at MPI@LHC'08, ``Multiple Partonic Interactions at the LHC'',  
Perugia, Italy, October 2008.}
} 
\author{Matteo Cacciari $^{1,2}$
}
\institute{$^1$LPTHE, UPMC -- Paris 6, 
CNRS UMR 7589, Paris, France\\
$^2$Universit\'e  Paris-Diderot -- Paris 7, Paris, France}
\maketitle
\begin{abstract}
Recent developments in jet clustering are reviewed. We present a list
of fast and infrared and collinear safe algorithms, and also describe new 
tools like jet areas. We show how these techniques can be applied to
the study of underlying event or, more generally, of any background which can be
considered distributed in a sufficiently uniform way.

\end{abstract}

\section{Recent Developments in Jet Clustering}

The final state of a high energy hadronic collision is inherently extremely
complicated. Hundreds or even thousands of particles will be
recorded by detectors at the Large Hadron Collider (LHC), making the task of reconstructing the original
(simpler) hard event very difficult. This large number of particles is the product of
a number of branchings and decays which follow the initial production of a
handful of partons. Usually only a limited number of stages of this production process can
be meaningfully described in quantitative terms, for instance by perturbation
theory in QCD. This is why, in order to compare theory and data, the latter must
first be {\sl simplified} down to the level described by the theory.

Jet clustering algorithms offer precisely this possibility of creating calculable observables
from many final-state particles. This is done by clustering them into jets via a
well specified algorithm, which usually contains one or more 
parameters, the most important of them being a ``radius'' $R$
which controls the extension of the jet in the rapidity-azimuth plane. One
can also choose a recombination scheme, which controls how partons' (or jets')
four-momenta are combined. The choice of a {\sl jet algorithm}, its {\sl parameters}
and the {\sl recombination scheme} is called a {\sl jet definition} \cite{Buttar:2008jx}, 
and must be
specified in full (together with the initial particles sample) in order for the
process
\begin{equation}
\left\{\mathrm {particles}\right\} \stackrel{\mathrm{jet~definition}}{\longrightarrow}
\left\{\mathrm {jets}\right\}
\end{equation}
to be fully reproducible and the final jets to be the same.

While (almost) any jet definition can produce sensible observables, not all of them
will produce one which is {\sl calculable} in perturbation theory. For this to
be true, the jet algorithm must be {\sl infrared and collinear safe} (IRC safe)
\cite{Sterman:1977wj},
meaning that actions producing configurations that lead to divergences in
perturbation theory, namely the emission of a very soft particle or a collinear splitting
of a particle into two) must not produce any change in the jets returned by the
algorithm.

The importance for jet algorithms to be IRC safe had been recognized as early as 1990 in the
`Snowmass accord' \cite{Huth:1990mi}, together with the need for them to be easily 
applicable both on the theoretical and the
experimental side. However, many of the implementations of jet clustering algorithms  used in the
following decade and a half failed to provide these characteristics: cone-type algorithms were
typically infrared or collinear unsafe beyond the two or three particle level (see
\cite{Buttar:2008jx} for a review), whereas recombination-type  algorithms were usually considered
too slow to be usable at the experimental level in hadronic collisions.

This deadlock was finally  broken by two papers, one in in 2005 \cite{Cacciari:2005hq}, which made
sequential recombination type clustering algorithms like $k_t$ \cite{Catani:1993hr,Ellis:1993tq} and
Cambridge/Aachen \cite{Dokshitzer:1997in, Wobisch:1998wt} fast, and one in 2007, which introduced
SISCone \cite{Salam:2007xv}, a cone-type algorithm which is infrared and collinear safe. A
third paper introduced, in 2008, the anti-$k_t$ algorithm \cite{Cacciari:2008gp}, a fast, IRC safe
recombination-type algorithm which however behaves, for many practical purposes, like a
nearly-perfect cone. This set of algorithms (see Table~\ref{table:safe-algs}), all available
through the {\tt FastJet} package \cite{fastjet_web},  allows one to replace most of the unsafe
algorithms still in use with fast and IRC safe ones, while retaining their main characteristics
(for instance, the MidPoint and the ATLAS cone could be replaced by SISCone, and the CMS cone
could be replaced by anti-$k_t$).

\begin{table}[t]
\begin{center}
\begin{tabular}{|c|c|c|}
\hline
Jet algorithm    & Type of algorithm, (distance measure)                         & algorithmic complexity \\
\hline
$k_t$   \protect\cite{Catani:1993hr,Ellis:1993tq}        & SR, $d_{ij} = \min(k_{ti}^2,k_{tj}^2) \Delta R_{ij}^2/R^2$       & $N\ln N$  \\
Cambridge/Aachen  \protect\cite{Dokshitzer:1997in, Wobisch:1998wt}& SR, $d_{ij} = \Delta R_{ij}^2/R^2$                              & $N\ln N$  \\
anti-$k_t$   \protect\cite{Cacciari:2008gp}     & SR, $d_{ij} = \min(k_{ti}^{-2},k_{tj}^{-2}) \Delta R_{ij}^2/R^2$ & $N^{3/2}$ \\
SISCone   \protect\cite{Salam:2007xv}        & seedless iterative cone with split-merge                       & $N^2\ln N$\\
\hline
\end{tabular}
\caption{\label{table:safe-algs} List of some of the IRC safe algorithms available in {\tt FastJet}. SR stands for
`sequential recombination'. $k_{ti}$ is a transverse momentum, and the angular distance is 
given by $\Delta R_{ij}^2 = \Delta y_{ij}^2 + \Delta \phi_{ij}^2$.}
\end{center}
\end{table}

\section{Jet Areas}

A by-product of the speed and the infrared safety of the new algorithms (or new
implementations of older algorithms) was found to be the possibility to define in a
practical way the {\sl area} of a jet, which measures its susceptibility to be
contaminated by a uniformly distributed background of soft particles in a given event.

\begin{figure}[t]
\begin{center}
\includegraphics[width=0.43\textwidth]{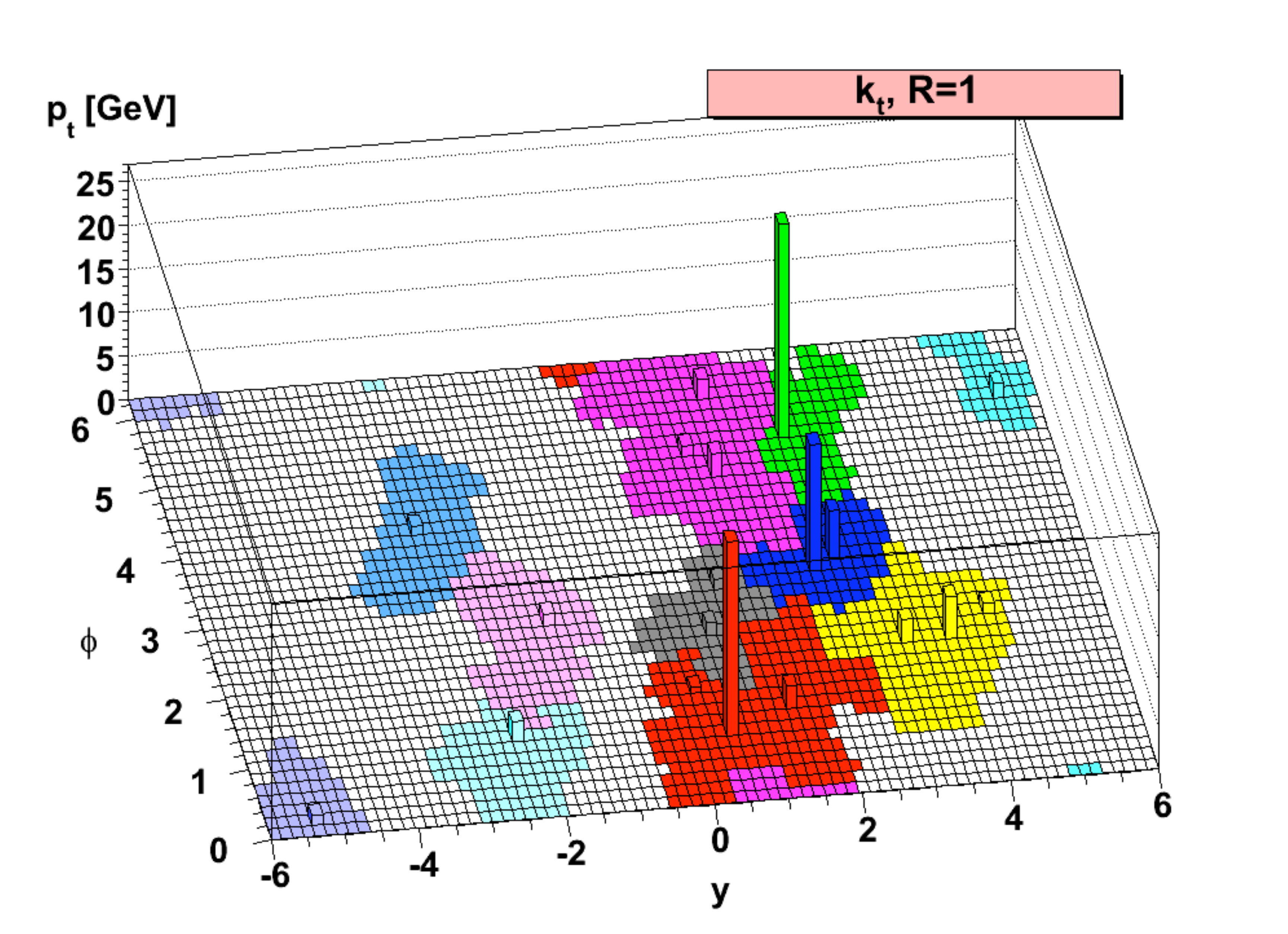}
~~~~~\includegraphics[width=0.43\textwidth]{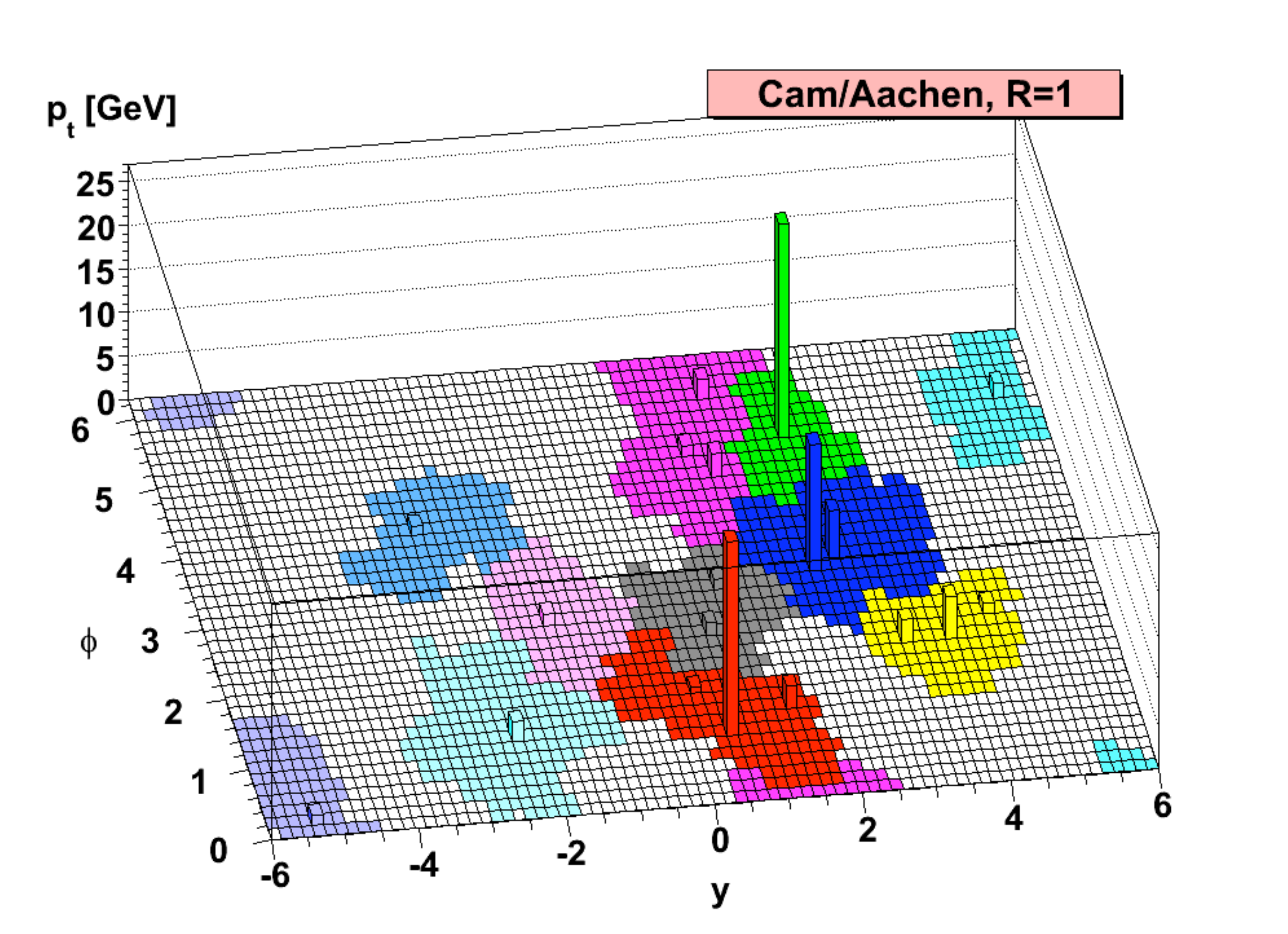}
\includegraphics[width=0.43\textwidth]{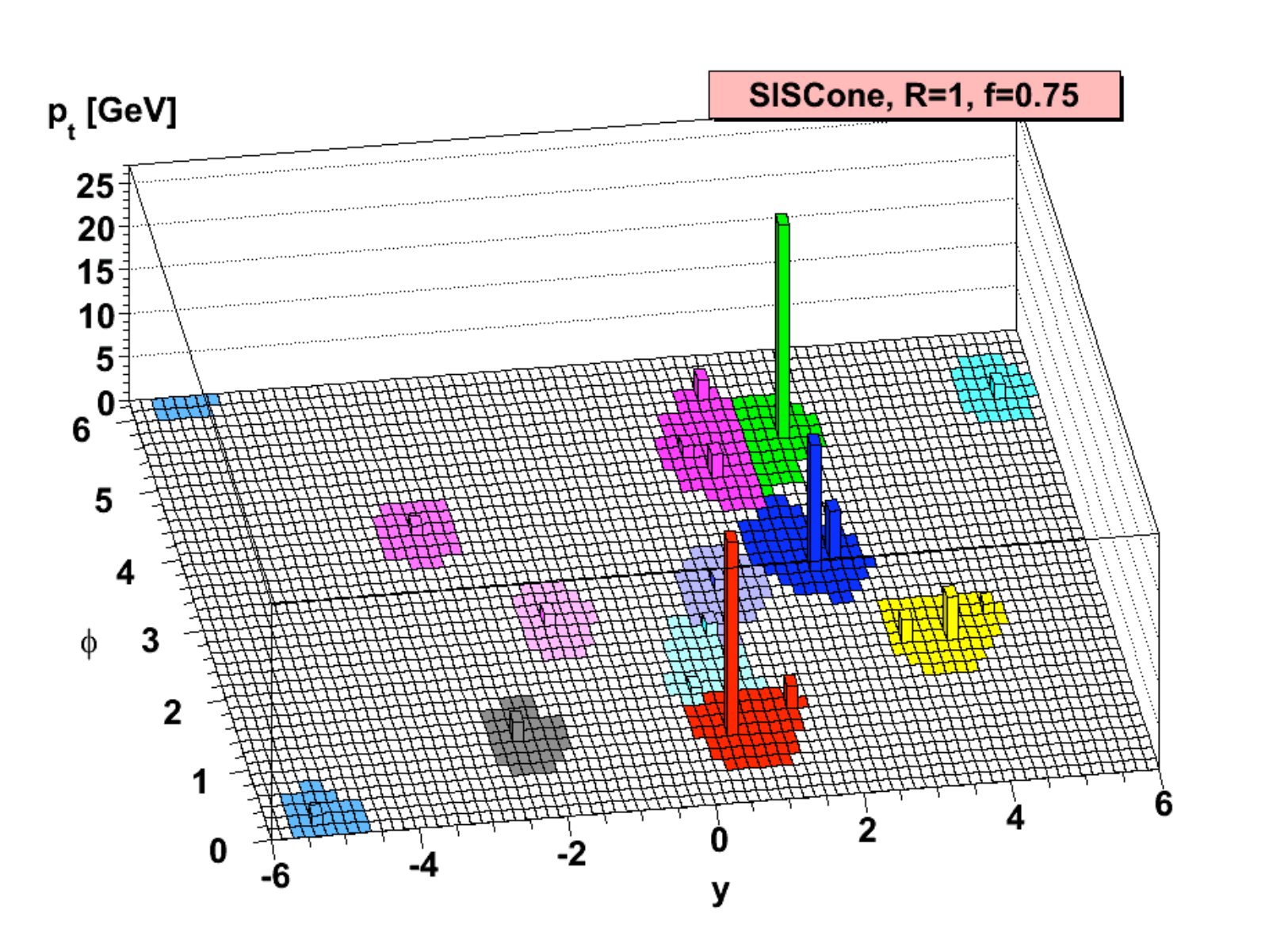}
~~~~~\includegraphics[width=0.43\textwidth]{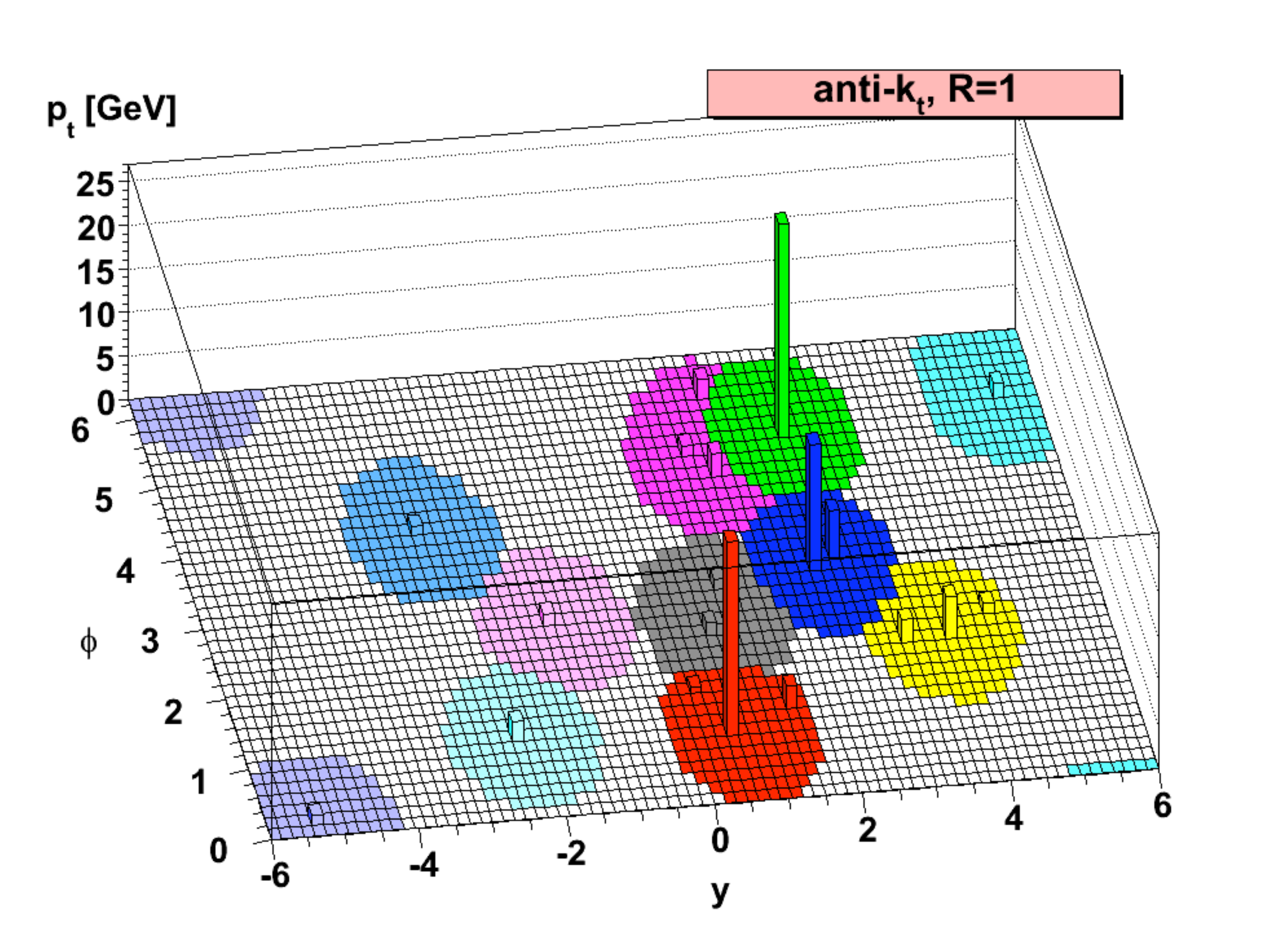}
\caption{\label{fig:areas} Typical jet outlines returned by four different IRC safe jet clustering
algorithms. From\protect\cite{Cacciari:2008gp}.}
\end{center}
\end{figure}

In their most modest incarnation, jet areas can be used to visualize the outline of the
jets returned by an algorithm so as to appreciate, for instance, if it returns regular
(``conical'') jets or rather ragged ones. An example is given in Fig.~\ref{fig:areas}.

Jet areas are amenable, to some extent, to analytic treatments \cite{Cacciari:2008gn}, or can be
measured numerically with the tools provided by {\tt FastJet}. These analyses disprove the common
assumption that all cone-type algorithms have areas equal to $\pi R^2$. In fact,
depending on exactly which type of cone algorithm one considers, its areas can differ, even
substantially so, from this naive estimate: for instance, the area of a SISCone jet made of a
single hard particle immersed in a background of many soft particles is $\pi
R^2/4$ (this
little catchment area can explain why other iterative cone algorithms with a split-merge step,
like the MidPoint algorithm in use at CDF, have often been seen to fare `well' in
noisy environments). One can analyse next the $k_t$ and the Cambridge/Aachen algorithms, and see
that their
single-hard-particle areas turn out to be roughly $0.81\pi R^2$. Finally, this area for the
anti-$k_t$ algorithm is instead exactly $\pi R^2$. This fact, together with its regular contours
shown in Fig.~\ref{fig:areas}, explains why it is usually considered to behave like a `perfect
cone'.

Jet areas also allow one to use some jet algorithms as tools to 
measure the level of a sufficiently uniform background which accompanies harder events.
This can be accomplished by following the procedure outlined in \cite{Cacciari:2007fd}: for each
event, all particles are clustered into jets using either the $k_t$ or the Cambridge/Aachen
algorithms, and the transverse momentum $p_{t,j}$ and the area $A_j$ of each jet are calculated. 
One observes that
a few hard jets have large values of transverse momentum
divided by  area, whereas most of the other, softer jets have smaller (and similar) values of this
ratio.
The background level $\rho$, transverse momentum per unit area in the
rapidity-azimuth plane, is then obtained as
\begin{equation}
\rho = \mathrm{median}\left\{\frac{p_{t,j}}{A_{j}}\right\}_{j \in {\cal R}} \, .
\end{equation}
The range ${\cal R}$ should be the largest possible 
region of the rapidity-azimuth plane over which the background is expected to be constant. 

The operation of taking the median of the $\{p_{t,j}/A_{j}\}$ distribution is,
to some extent, arbitrary. It has been found to give  sensible results,
provided that the range ${\cal R}$ contains sufficiently many soft background jets -- at
least about ten (twenty) of them, if only one (two) harder jets are also present in ${\cal R}$, 
are usually enough \cite{css}.

\section{Underlying Event Studies}

To a certain extent, and within certain limits, the background to a hard collision created by the
soft particles of the underlying event (EU)  can be considered fairly uniform. It becomes then
amenable to be studied with the technique introduced in the previous Section. 
This constitutes an alternative to the usual and widespread approach (see for
instance \cite{Acosta:2004wqa,Acosta:2006bp}) of triggering on a
leading jet, and selecting the two regions in the azimuth space which are transverse to its
direction and to that of the recoil jet. These two regions are considered to be little
affected by hard radiation (in the least energetic of them it is expected to be suppressed
by at least two powers of $\alpha_s$), and therefore one can expect to be able 
to measure the UE level there.

\begin{figure}[t]
\begin{center}
\includegraphics[width=0.7\textwidth]{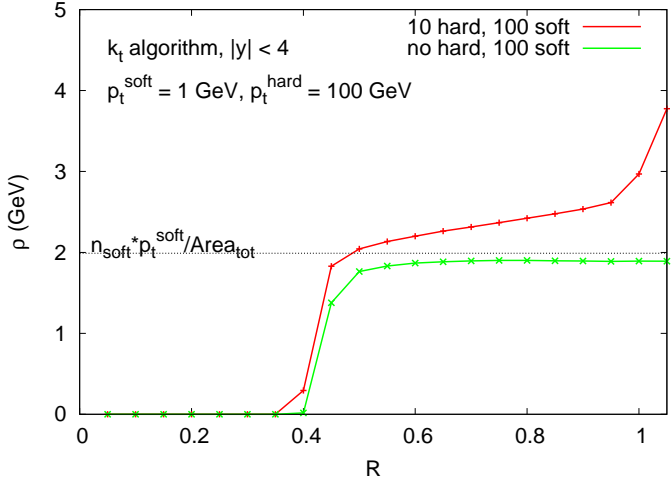}
\caption{\label{fig:UE} Determination of the background level $\rho$ of a toy-model random 
underlying event,
as a function of the radius parameter $R$. Each point is the result of averaging over many
different realizations. The parameters have been adjusted to roughly reproduce the situation
expected at the LHC.}
\end{center}
\end{figure}

This way of selecting the UE can be considered a {\sl topological} one: particles (or jets) are
classified as belonging to the UE or not as a result of their position. On the other hand, the
median procedure described in the previous Section can be thought of as a  {\sl dynamical
selection}: no a priori hypotheses are made and, in a way that changes from one event to another, 
a jet is automatically classified as belonging to
the hard event or to the background as a result of its characteristics (namely the value of the
$p_{t,j}/A_{j}$ ratio). One can further show that this selection pushes the possible
contamination from perturbative radiation to very large powers of $\alpha_s$: for a range ${\cal
R}$ defined by $|y| < y_{max}$, perturbative contamination will only start at order  $n \simeq 3
y_{max}/R^2$ \cite{Cacciari:2007fd}. This gives $n \sim 24$ for $y_{max} = 2$ and $R=0.5$,
suggesting that the perturbative contribution is minimal.

A sensible criticism of this procedure is that the UE distribution is not necessarily
uniform, and may for instance vary as a function of rapidity. A way around this is then to
choose smaller ranges, located at different rapidity values, and repeat the $\rho$
determination in each of them. Of course care will have to be taken that the chosen ranges
remain large enough to satisfy the criterion on the number of soft jets versus hard ones
given in the previous Section: for instance, a range one unit of rapidity large can be
expected to contain roughly $2\pi/(0.55 \pi R^2) \sim 15$ soft jets for $R=0.5$, which
makes it marginally apt to the task\footnote{Its performance can be improved by removing the
hardest jets it contains from the $\{p_{t,j}/A_{j}\}$ list before taking the median \cite{css}.}.

A final word should be spent on which values of the radius parameter $R$ can be considered
appropriate for this analysis. Roughly speaking, $R$ should be large enough for the number of
`real' jets (i.e. containing real particles) to be at last larger than the number of `empty jets'
(regions of the rapidity-azimuth plane void of particles, and not occupied by any `real' jet). It
should also be small enough to avoid having too many jets containing too many hard particles.
Analytical estimates \cite{Cacciari:2007fd} and empirical evidence show that for UE estimation in
typical LHC conditions one can expect values of the order of 0.5 -- 0.6 to be appropriate. Much
smaller values will return $\rho \simeq 0$, while larger values will tend to return progressively
larger values of $\rho$, as a result of the increasing contamination from the hard jets. Fig.
\ref{fig:UE} shows results obtained with a toy model where 100 soft particles with $p_T^{soft}
\simeq 1$ GeV are generated in a $|y|<4$ region. Ten hard particles, with $p_T^{hard} \simeq 100$
GeV, can be additionally generated in the same region. One observes how, after a threshold value for
$R$, $\rho$
is estimated correctly for the soft-only case, while when hard particles are present they
increasingly contaminate the estimate of the background.

\begin{figure}[t]
\begin{center}
\includegraphics[width=0.45\textwidth]{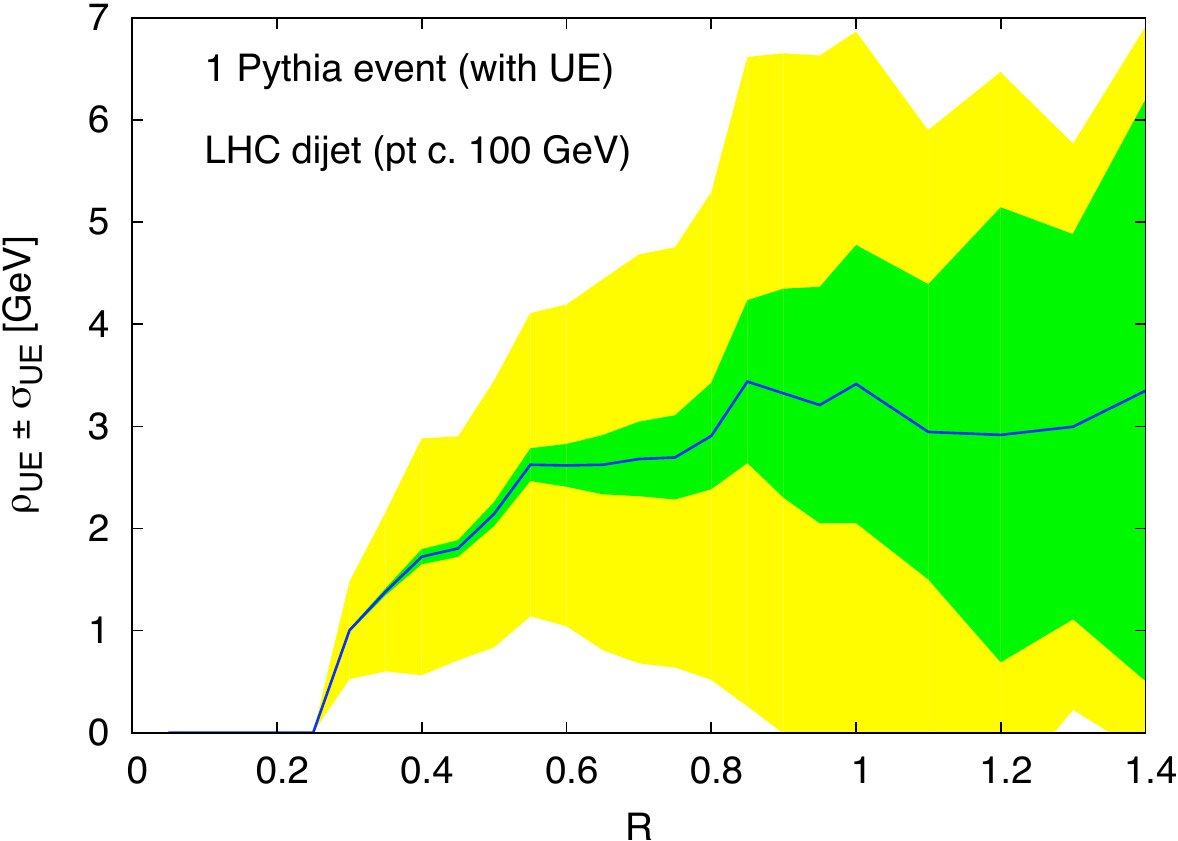}
~~~~~~
\includegraphics[width=0.45\textwidth]{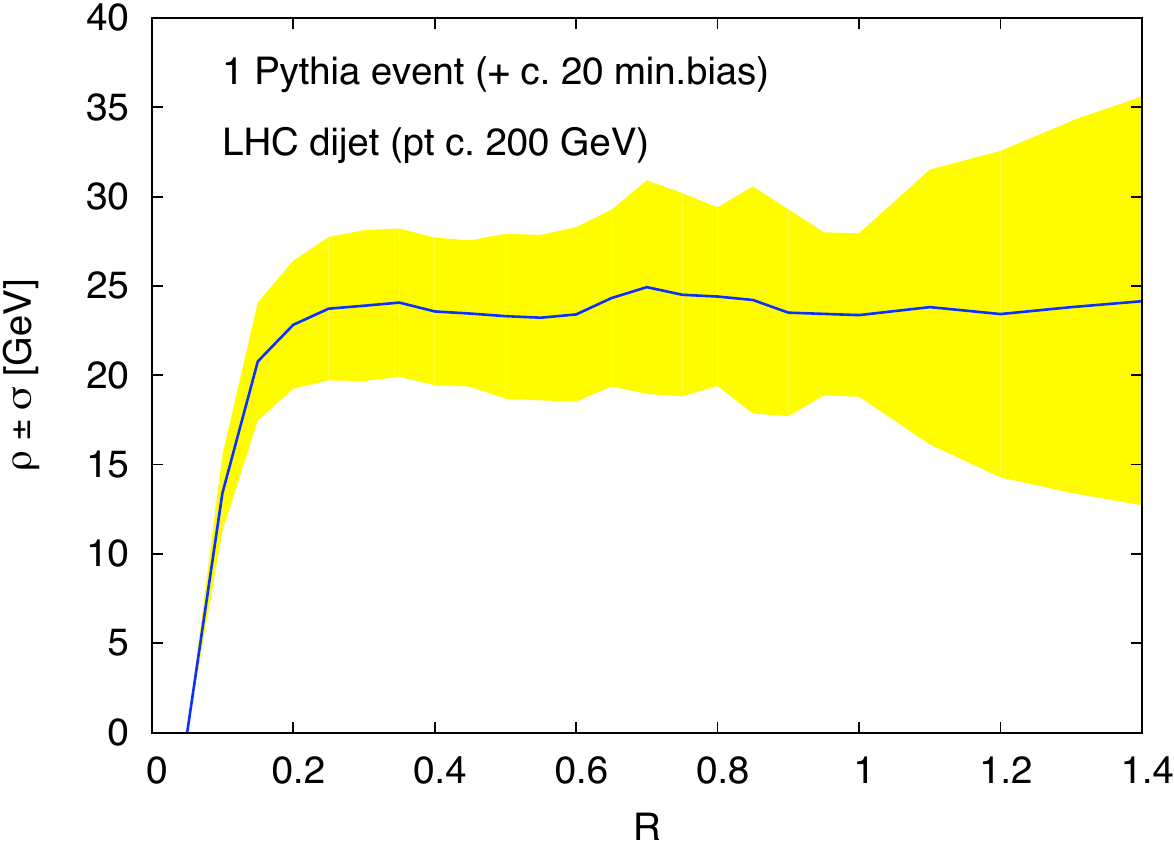}
\caption{\label{fig:pu-ue} Determination of the background level $\rho$ in realistic dijet
events at the LHC, with (right)  and without (left) pileup. Preliminary results.}
\end{center}
\end{figure}

The same analysis can be performed on more realistic events, generated by Monte Carlo simulations.
Fig. \ref{fig:pu-ue} shows the determination of $\rho$ in a simulated dijet event at the LHC, with
and without pileup. In both cases the general structure of the toy-model in Fig. \ref{fig:UE} can
be seen, though it is worth noting that in the UE case (left plot) the slope can vary significantly
from event to event, and also according to the Monte Carlo tune used \cite{css}. 
The larger particle density
(and probably higher uniformity) of the pileup case allows for an easier and more stable
determination.

\begin{figure}[t]
\begin{center}
\includegraphics[width=0.8\textwidth]{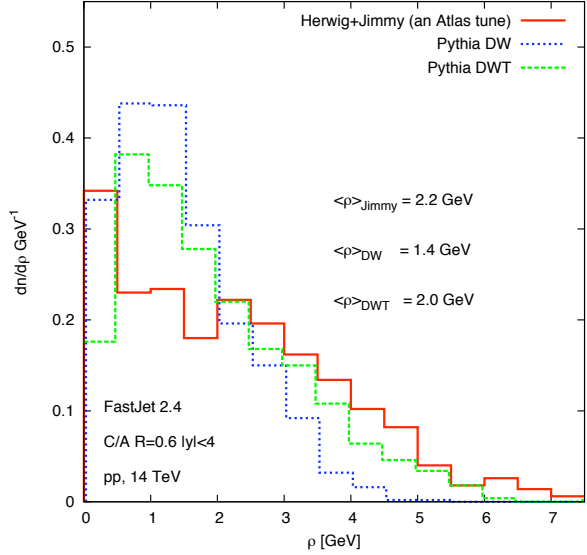}
\caption{\label{fig:css} Distributions  of $\rho$ from the UE over many simulated LHC dijet events ($p_T > 50$ GeV, $|y| < 4$),
using different Monte Carlos and different UE tunes. Preliminary results.}
\end{center}
\end{figure}

Once a procedure for determining $\rho$ is available, one can think of many different
applications. One possibility is of course to tune Monte Carlo models to real data by comparing rho
distributions, correlations, etc. A preliminary example is given in fig. \ref{fig:css}, where
studying the distribution of $\rho$ can be seen to allow one to discriminate between UE models
which would otherwise give similar values for the average contribution $\langle\rho\rangle$.
More extensive studies are in progress \cite{css}.

Yet another use of measured $\rho$ values is the {\sl subtraction} of the background from the
transverse momentum of hard jets. Ref. \cite{Cacciari:2007fd} proposed to correct the four-momentum
$p_{\mu j}$ of the jet $j$ by
an amount proportional to $\rho$ and to the area of the jet itself (the susceptibility of the jet
to contamination):
\begin{equation}
p_{\mu j}^{sub} = p_{\mu j} - \rho A_{\mu j}
\end{equation}
where $A_{\mu j}$ is a four-dimensional generalization of the concept of jet area, normalized in
such a way that its transverse component coincides, for small jets, with the scalar area $A_j$
\cite{Cacciari:2008gn}. One can show \cite{Cacciari:2007fd,Cacciari:2008gd} that such subtraction of the underlying event
can improve in a non-negligible way the reconstruction of mass peaks even at very large energy
scales. A similar procedure is also being considered \cite{crss} 
for heavy ion collisions,
where the background can contribute a huge contamination, even larger
than the transverse momentum of the hard jet itself (partly because of this, one usually speaks 
of `jet
reconstruction' in this context, rather than just `subtraction'). 
Initial versions of this technique have
already been employed at the experimental level by the STAR Collaboration at
RHIC in \cite{Salur:2008hs,Salur:2008vm}, where IRC safe jets have been
reconstructed for the first time in heavy ion collisions.

\section{Conclusions}

Since 2005 numerous developments have intervened in jet physics. A number of fast and infrared and
collinear safe algorithms are now available, allowing for great flexibility in analyses. Tools have
been developed and practically implemented to calculate jet areas, and these can used to study
various types of backgrounds (underlying event, pileup, heavy ions background) and also to
subtract their contribution to large transverse-momentum jets.

These new algorithms and methods (as well as the ones not mentioned in this talk, like the many
approaches to jet substructure, see e.g. 
\cite{Butterworth:2008iy,Thaler:2008ju,Kaplan:2008ie,Almeida:2008yp,Ellis:2009su}, useful in
a number of new-physics searches) are transforming jet physics from being just a procedure to obtain 
calculable observables
to providing a full array of precision tools with which to probe efficiently the complex final
states of high energy collisions.

\section*{Acknowledgments}

I wish to thank the organizers of MPI@LHC'08 in Perugia for the invitation to this interesting
conference, as well as Gavin P. Salam and Sebastian
Sapeta for collaboration on the ongoing underlying event studies, and Gregory Soyez and 
Juan Rojo for the work done together on related jet issues.

\begin{footnotesize}
\bibliographystyle{mpi08} 
{\raggedright
\bibliography{proc}
}
\end{footnotesize}
\end{document}